 \documentclass[12pt]{article}
  \usepackage{epsfig}
\newcommand{\inc}{{\it i}}

 \newcommand{\be}{\begin{equation}}
 \newcommand{\ee}{\end{equation}}
 \newcommand{\ba}{\begin{eqnarray}}
 \newcommand{\ea}{\end{eqnarray}}
 \newcommand{\efbold}{\mbox{{\boldmath $\vec f$}}}
 
 \newcommand{\erbold}{\mbox{{\boldmath $\vec r$}}}
 \newcommand{\Phibold}{\mbox{{\boldmath $\vec \Phi$}}}

 \newcommand{\pbold}{\mbox{{\boldmath $\vec p$}}}

 \newcommand{\doterbold}{\dot{\textbf {\mbox{\boldmath $\vec {\boldmath r}$}} }}

\begin{document}
% \date{}
\title{
% ${~}^{{Letter~to~the~Editor}}$~~~~~~~~~~~~~~~~~~${~}^{\left.~~~\right.\underline{To~be~submitted~to~\it{A\&A}}}$
%
%
 ${}^{Talk~at~the~conference~``Journ\acute{e}es~2004:~Syst\grave{e}mes~de~r\acute{e}f\acute{e}rence~spatio-temporels,"}$\\
 ${}^{l'Observatoire~de~Paris,~20~-~22~septembre~2004.}$
\\
 {\Large{\textbf{
  On the theory of canonical perturbations and its applica- tion to Earth rotation.
 A source of inaccuracy in the calculation of the angular velocity
% On the theory of canonical perturbations and its (mis)application to Earth rotation
 }}}\\ }
 \author{ {\Large{Michael Efroimsky}}\\ {\small{US Naval Observatory,
 Washington DC 20392 USA}}, {~~~\small{e-mail: me @ usno.navy.mil~}}
  }
  \maketitle
 \begin{abstract}

Both orbital and rotational dynamics employ the method of
variation of parameters. We express, in a non-perturbed setting,
the coordinates (Cartesian, in the orbital case, or Eulerian in
the rotation case) via the time and six adjustable constants
called elements (orbital elements or rotational elements). If,
under disturbance, we use this expression as ansatz and endow the
``constants" with time dependence, then the perturbed velocity
(Cartesian or angular) will consist of a partial derivative with
respect to time and a so-called convective term, one that includes
the time derivatives of the variable ``constants." Out of sheer
convenience, the so-called Lagrange constraint is often imposed.
It nullifies the convective term and, thereby, guarantees that the
functional dependence of the velocity upon the time and
``constants" stays, under perturbation, the same as it used to be
in the undisturbed setting. The variable ``constants" obeying this
condition are called osculating elements. Otherwise, they are
simply called orbital or rotational elements.

When the dynamical equations, written in terms of the
``constants," are demanded to be symplectic (and the ``constants"
make conjugated pairs $\,Q,\,P$), these ``constants" are called
Delaunay elements, in the orbital case, or Andoyer elements, in
the rotational case. The Andoyer and Delaunay sets of elements
share a feature not readily apparent: in certain cases, the
standard equations render these elements non-osculating.

In orbital mechanics, the elements, calculated via the standard
planetary equations, come out non-osculating when perturbations
depend on velocities. This complication often arises but seldom
gets noticed. To keep elements osculating under such
perturbations, extra terms must enter the equations, terms that
will {\textbf{not}} be parts of the disturbing function (Efroimsky
\& Goldreich 2003, 2004). In the case of parametrisation through
the Kepler elements, this will merely complicate the equations. In
the case of Delaunay parametrisation, these extra terms will not
only complicate the Delaunay equations, but will also destroy
their canonicity. Under velocity-dependent disturbances, the
osculation and canonicity conditions are incompatible.

Similarly, in rotational dynamics, the Andoyer elements come out
non-osculating when the perturbation depends upon the angular
velocity of the top. Since a switch to a non-inertial frame is an
angular-velocity-dependent perturbation, then amendment of the
dynamical equations by only adding extra terms to the Hamiltonian
makes these equations render non-osculating Andoyer elements. To
make them osculating, extra terms must be added to the dynamical
equations (and then these equations will no longer be symplectic).

Calculations in terms of non-osculating variables are
mathematically valid, but their physical interpretation is
problematic. Non-osculating orbital elements parametrise
instantaneous conics \textbf{not} tangent to the orbit. Their
inclination, the non-osculating {\it i}, may differ much from the
physical inclination of the orbit, given by the osculating
{\inc}$\,$. Similarly, in the case of rotation, non-osculating
Andoyer variables do correctly describe a perturbed spin but lack
simple physical meaning. The customary expressions for the
spin-axis' orientation, in terms of the Andoyer elements, will no
longer be valid, if the elements are non-osculating. These
expressions, though, will stay valid for osculating elements, but
then the (correct) dynamical equations for such elements will no
longer be canonical -- circumstance ignored in the
Kinoshita-Souchay (KS) theory which tacitly employs non-osculating
variables. While the loss of osculation will not influence the
predictions for the figure axis of the planet, it considerably
effects the predictions for the orientation of the instantaneous
axis of rotation.
%\
%\\
% KEY WORDS: ~~~~Orbit integration, Lagrange system, Delaunay system,\\
% {${\left. \; \; \right.}^{\left. \; \right.}\;$}~~~~~~~~~~~~~~~~~~~~~Hidden symmetries, Gauge invariance.\\
% RUNNING HEAD: ~Hidden symmetry of the Lagrange and Delaunay systems
\end{abstract}

\section{Kepler and Euler}

In orbital dynamics, a Keplerian conic, emerging as an undisturbed
two-body orbit, is regarded as a sort of ``elementary motion," so
that all the other available motions are conveniently considered
as distortions of such conics, distortions implemented through
endowing the orbital constants $\;C_j\;$ with their own time
dependence. Points of the orbit can be contributed by the
``elementary curves" either in a non-osculating fashion, as in
Fig. 1, or in the osculating way, as in Fig. 2.

 The disturbances, causing the evolution of the motion from one
 instantaneous conic to another, are the primary's oblateness,
 the gravitational pull of other bodies, the atmospheric and
 radiation-caused drag, and the non-inertiality of the reference system.

Similarly, in rotational dynamics, a complex spin can be presented
as a sequence of configurations borrowed from a family of some
elementary rotations. The easiest possibility here will be to
employ in this role the Eulerian cones, i.e., the loci of the
rotational axis, corresponding to non-perturbed spin states. These
are the simple motions exhibited by an undeformable free top with
no torques acting thereupon.\footnote{~Here one opportunity will
be to employ in the role of ``elementary" motions the non-circular
Eulerian cones described by the actual triaxial top, when this top
is unforced. Another opportunity will be to use, as ``elementary"
motions, the circular Eulerian cones described by a dynamically
symmetrical top (and to treat its actual triaxiality as another
perturbation). The main result of our paper will be invariant
under this choice.} Then, to implement a perturbed motion, we
shall have to go from one Eulerian cone to another, just as in
Fig. 1 and 2 we go from one Keplerian ellipse to another. Hence,
similar to those pictures, a smooth ``walk" over the instantaneous
Eulerian cones may be osculating or non-osculating.

The physical torques, the actual triaxiality of the top, and the
non-inertial nature of the reference frame will then be regarded
as perturbations causing the ``walk." The latter two perturbations
depend not only upon the rotator's orientation but also upon its
angular velocity.

\section{Delaunay and Andoyer}

In orbital dynamics, we can express the Lagrangian of the reduced
two-body problem via the spherical coordinates
$\;q_j\;=\;\{\,r\,,\;\varphi\,,\;\theta\,\}\;$, then calculate
their conjugated momenta $\;p_j\;$ and the Hamiltonian $\;{\cal
H}(q,\,p)\;$, and then carry out the Hamilton-Jacobi procedure
(Plummer 1918), to arrive to the Delaunay variables
 \ba
% $\;
 \nonumber
\{\,Q_1\,,\;Q_2\,,\;Q_3\;;\;P_1\,,\;P_2\,,\;P_3\,\}\,\equiv\,
\{\,L\,,\;G\,,\;H\;;\;{\it l}\,,\;g\,,\;h\,\}\;=~~~~~~~~~\\
\label{1}\\
 \nonumber
\{\,\sqrt{\mu a}\;\;,\;\;\sqrt{\mu a \left(1\,-\,e^2
\right)}\;\;,\;\;\sqrt{\mu a \left(1\,-\,e^2 \right)}\;\cos
\inc\;\;;\;\;-\,M_o\;\;,\;\;-\,\omega\;\;,\;\;-\,\Omega \,\} \;,
% $
\ea
 where $\;\mu\;$ denotes the reduced mass.

 Similarly, in rotational dynamics one can define a spin
state of a top by means of the three Euler angles
$\,q_j\,=\,\psi\,,\,\theta\,,\,\varphi\,$ and their canonical
momenta $\,p_j\,$, and then perform a canonical transformation to
the Andoyer elements $\,L\,,\,G\,,\,H\,,\,{\it l}\,,\,g\,,\,h\,$.
A minor technicality is that, historically, these variables were
introduced by Andoyer (1923) in a manner slightly different from
the set of canonical constants: while, for a free rotator, the
three Andoyer variables $\,G\,,\,H\,,\,h\,$ are constants, the
other three, $\,L\,,\,{\it l}\,,\,g\,$ do evolve in time (for the
Andoyer Hamiltonian of a free top is not zero,
 \begin{center}
        \epsfxsize=34mm
        \epsfbox{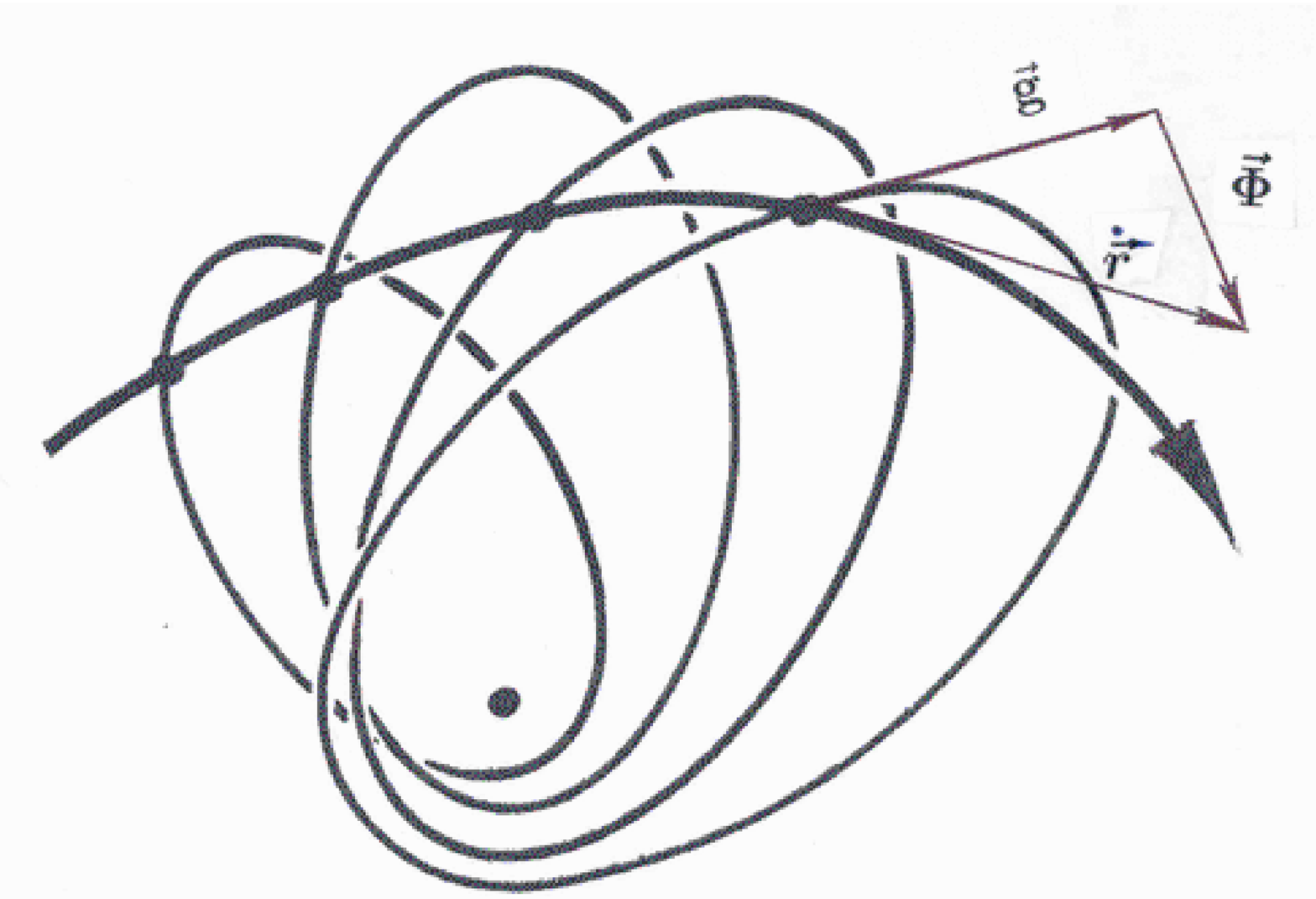}
%        _corrected.eps}
  \end{center}
  \mbox{\small
 \parbox[b]{5.5in}{{\underline{Fig.1.}} ~~~ \small The perturbed trajectory
 is a set of points belonging to a sequence of
confocal instantaneous ellipses. The ellipses are \textbf{not}
supposed to be tangent, nor even coplanar to the orbit at the
intersection point. As a result, the physical velocity
$\,\doterbold\,$ (tangent to the trajectory at each of its points)
differs from the Keplerian velocity $\,\bf\vec g\,$ (tangent to
the ellipse). To parametrise the depicted sequence of
non-osculating ellipses, and to single it out of all the other
such sequences, it is suitable to employ the difference between
$\,\doterbold\,$ and $\,\bf\vec g\,$, expressed as a function of
time and six (non-osculating) orbital elements:\\
% \ba
% \nonumber
 $\;
 \Phibold(t\,,\;C_1\,,\;.\,.\,.\,,\;C_6)\;=\;
 \doterbold(t\,,\;C_1\,,\;.\,.\,.\,,\;C_6)\,-\,
 {\bf\vec g}(t\,,\;C_1\,,\;.\,.\,.\,,\;C_6)\;\;.\;\;
 $\\
% \ea
 Since
 \ba
 \nonumber
 \doterbold\,=\,\frac{\partial \erbold}{\partial
 t}\,+\,\sum_{j=1}^{6}\frac{\partial C_j}{\partial
 t}\;\dot{C}_j\;=\;{\bf\vec g}\;+\;\sum_{j=1}^{6}\frac{\partial C_j}{\partial
 t}\;\dot{C}_j\;\;\;,
 \ea
 then the difference $\,\Phibold\,$ is simply the
 convective term $\;\sum \left(\partial \erbold/\partial C_j
 \right)\,\dot{C}_j\;$ which emerges whenever the
 instantaneous ellipses are being gradually altered by the
 perturbation (and the orbital elements become time-dependent).
 In the literature,  $\;\Phibold(t\,,\;C_1\,,\;.\,.\,.\,,\;C_6)\;$ is
 called the
 gauge function or gauge velocity or, simply, gauge.
 }}

 ~\\
 \begin{center}
        \epsfxsize=34mm
        \epsfbox{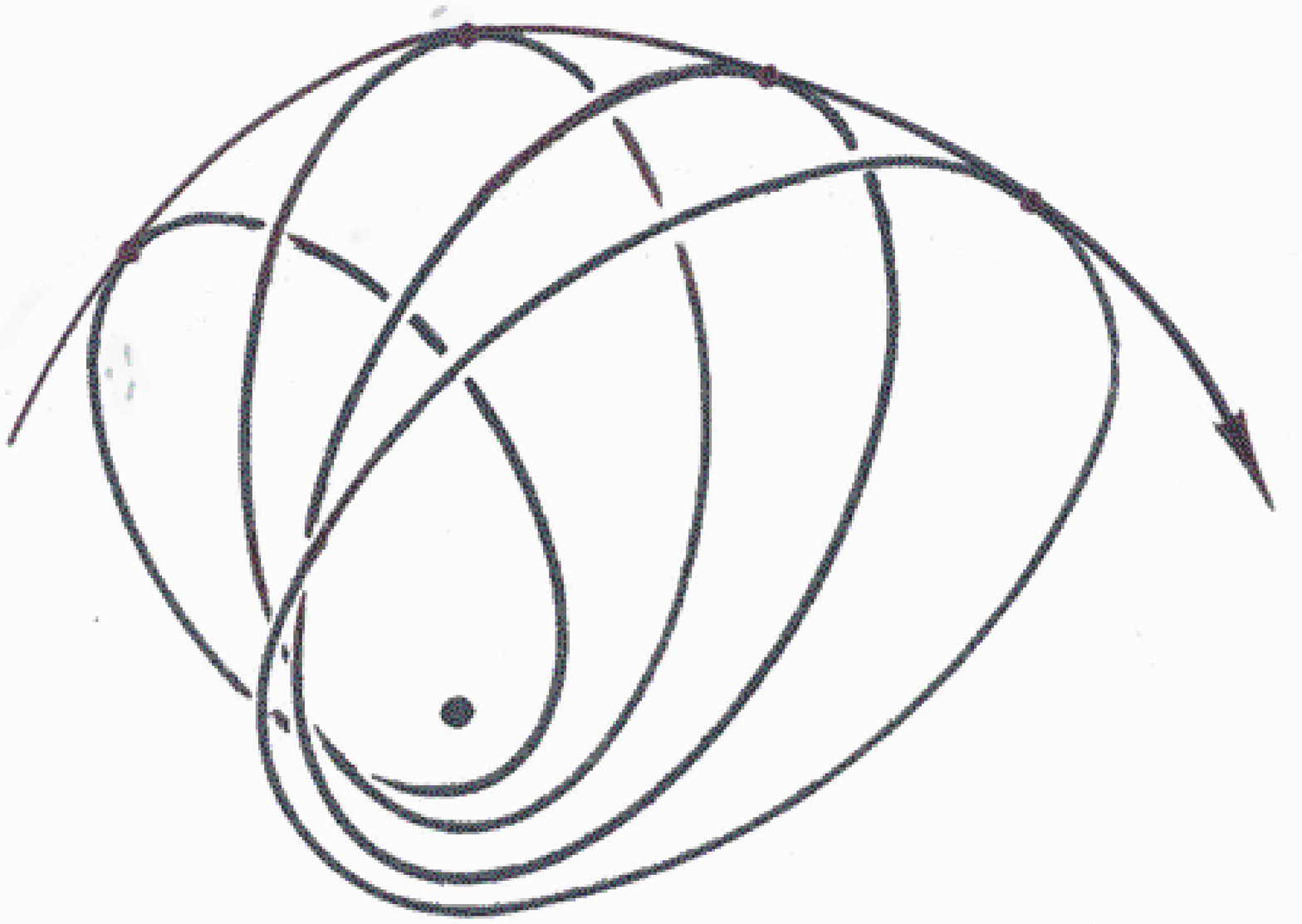}
 \end{center}
  \mbox{\small
 \parbox[b]{5.5in}{{\underline{Fig.2.}}~~~The perturbed trajectory
 is represented through a sequence of confocal instantaneous ellipses
 which {are} tangent to the trajectory at the
 intersection points, i.e., are osculating. Now, the physical velocity
$\,\doterbold\,$ (which is tangent to the trajectory) will
coincide with the Keplerian velocity $\,\bf\vec g\,$ (which is
tangent to the ellipse), so that their difference
$\;\Phibold(t\,\;C_1\,,\;.\,.\,.\,,\;C_6)\;$ vanishes everywhere:
 \ba
 \nonumber
 \Phibold(t\,,\;C_1\,,\;.\,.\,.\,,\;C_6)\;\equiv\;
 \doterbold(t\,,\;C_1\,,\;.\,.\,.\,,\;C_6)\,-\,
 {\bf\vec g}(t\,,\;C_1\,,\;.\,.\,.\,,\;C_6)\;=\;\sum_{j=1}^{6}\frac{\partial C_j}{\partial
 t}\;\dot{C}_j\;=\;0\;\;\;.
 \ea
 This equality, called Lagrange constraint or Lagrange gauge, is the necessary
 and sufficient condition of osculation.
}}

 \pagebreak

 \noindent
but a function of $\,{\it l}\,,\,L\,$ and $\,G\,$). This way, to
make our analogy complete, we may carry out one more canonical
transformation, from the Andoyer variables
$\,\{\,L\,,\,G\,,\,H\,,\,{\it l}\,,\,g\,,\,\,h\,\}\,$ to ``almost
Andoyer" variables $\,\{\,L_o\,,\,G\,,\,H\,,\,{\it
l}_o\,,\,g_o\,,\,h\,\}\,$, where $\,L_o\,,\,{\it l}_o\,$ and
$\,g_o\,$ are the initial values of $\,L\,,\,{\it l}\,$ and
$\,g\,$. The latter set consists only of the constants of
integration; the corresponding Hamiltonian becomes nil. Therefore,
these constants are the true analogues of the Delaunay variables
(while the conventional Andoyer set is analogous to the Delaunay
set with $\,M\,$ used instead of $\,M_o\,$.). The main result
obtained below for the modified Andoyer set
$\,\{\,L_o\,,\,G\,,\,H\,,\,{\it l}_o\,,\,g_o\,,\,h\,\}\,$ can be
easily modified for the regular Andoyer set of variables
$\,\{\,L\,,\,G\,,\,H\,,\,{\it l}\,,\,g\,,\,h\,\}\,$. (Efroimsky
2005b)

To summarise this section, in both cases we start out with
  \ba
 \dot{q}\;=\;\frac{\partial {\cal H}^{(o)}}{\partial p}\;\;\;,\;\;\;\;\;~~
 \dot{p}\;=\;-\;\frac{\partial {\cal H}^{(o)}}{\partial q}
 ~~~.~~~~~~~~~~~~~~~~~~~~~~~~~~~~~~~
 \label{2}
 \ea
$q\;$ and $\;p\;$ being the coordinates and their conjugated
momenta, in the orbital case, or the Euler angles and their
momenta, in the rotation case. Then we switch, via a canonical
transformation
 \ba
 \nonumber
 q\;=\;f(Q\,,\;P\,,\;t)\;\;\;\\
 \label{3}\\
 \nonumber
 p\;=\;\chi(Q\,,\;P\,,\;t)\;,\;
 \ea
 to
 \ba
 \dot{Q}\;=\;
 \frac{\partial {\cal H}^*}{\partial P}\;=0\;\;\;,\;\;\;\;\;
 \dot{P}\;=\;-\;\frac{\partial {\cal H}^*}{\partial Q}\;=\;0
 \;\;\;,\;\;\;\;
 {\cal H}^*\;=\;0\;\;,
 \label{4}
 \ea
 where $\;Q\;$ and $\;P\;$ denote the set of Delaunay elements, in the
 orbital case, or the (modified, as explained above) Andoyer
 set $\;\{\,L_o\,,\;G\,,\;H\,,\;{\it l}_o\,,\;g_o\,,\;h\,\}\;$, in
 the case of rigid-body rotation.

This scheme relies on the fact that, for an unperturbed Keplerian
orbit (and, similarly, for an undisturbed Eulerian cone) its
six-constant parametrisation may be chosen so that:\\
~~ \textbf{\underline{1.}}~~~the parameters are constants and, at
the same time, are canonical variables $\,\{\,Q\,,\,P\,\}\,$ with
a zero Hamiltonian:
$\,{\cal H}^*(Q,\,P)\,=\,0\,$; \\
~~\textbf{\underline{2.}}~~$\,$for constant $\,Q\,$ and $\,P\,$,
the transformation equations (\ref{3}) are mathematically
equivalent to the dynamical equations (\ref{2}).

\section{When do the elements come out non-osculating?}

\subsection{General-type motion}

Under perturbation, the ``constants" $\,Q\,,\,P\,$ begin to evolve
so that, after their substitution into
 \ba
 \nonumber
 q\;=\;f\left(\,Q(t)\,,\;P(t)\,,\;t\,\right)\;\;\;\\
 \label{5}\\
 \nonumber
 p\;=\;\chi(\,Q(t)\,,\;P(t)\,,\;t\,)\;\;\;
 \ea
($f\,$ and $\,\chi\,$ being the same functions as in
(\ref{3})$\,$), the resulting motion obeys the disturbed equations
  \ba
 \dot{q}\;=\;\frac{\partial \left({\cal H}^{(o)}\,+\,\Delta {\cal H} \right)}{\partial p}\;\;\;,\;\;\;\;\;~~
 \dot{p}\;=\;-\;\frac{\partial \left({\cal H}^{(o)}\,+\,\Delta {\cal H} \right)}{\partial q}
 ~~~.~~~~~~~~~~~~~~~
 \label{6}
 \ea
We also  want our ``constants" $\;Q\;$ and $\;P\;$ to remain
canonical and to obey
  \ba
 \dot{Q}\;=\;\frac{\partial \left({\cal H}^*\,+\,\Delta {\cal H}^* \right)}{\partial P}\;\;\;,\;\;\;\;\;~~
 \dot{P}\;=\;-\;\frac{\partial \left({\cal H}^*\,+\,\Delta {\cal H}^* \right)}{\partial Q}
 ~~~~~~~~~~~~~~~~~~
 \label{7}
 \ea
 where
 \ba
 {\cal H}^*\,=\;0\;\;\;\;\mbox{and}\;\;\;\;\;\Delta {\cal
 H}^*\left(Q\,,\;P\,\;t\right)\;=\;\Delta {\cal
 H}\left(\,q(Q,P,t)\,,\;p(Q,P,t)\,,\;t\,\right)\;\;\;.
 \label{8}
 \ea
Above all, an optimist will expect that the perturbed ``constants"
$\,C_j\,\equiv\,Q_1\,,\,Q_2\,,\,Q_3\,,\,P_1\,,\,P_2\,,\,P_3\;$
(the Delaunay elements, in the orbital case, or the modified
Andoyer elements, in the rotation case) will remain osculating.
This means that the perturbed velocity will be expressed by the
same function of $\,C_j(t)\,$ and $\,t\,$ as the unperturbed one
used to. Let us check to what extent this optimism is justified.
The perturbed velocity reads
 \ba
 \dot{q}\;=\;g\;+\;\Phi ~~~~~~~~~~~~~~~~~~~~~~~~~~~~~~
 \label{9}
 \ea
where
 \ba
 g(C(t),\,t)\;\equiv\;\frac{\partial q(C(t),\,t)}{\partial t}
 \;\;~~~~~~~~~~~~
 \label{10}
 \ea
is the functional expression for the unperturbed velocity; and
 \ba
 \Phi(C(t),\,t)\;\equiv\;\sum_{j=1}^6\,\frac{\partial q(C(t),\,t)}{\partial
 C_j}\;\dot{C}_j(t)\;
 \label{11}
 \ea
is the convective term. Since we chose the ``constants" $\,C_j\,$
to make canonical pairs $\,(Q,\,P)\,$ obeying (\ref{7} - \ref{8}),
with vanishing $\,{\cal H}^*\,$, then insertion of (\ref{7}) into
(\ref{26}) will result in
 \ba
 \Phi\;=\;\sum_{n=1}^3\,\frac{\partial q}{\partial
 Q_n}\;\dot{Q}_n(t)\;+\;\sum_{n=1}^3\,\frac{\partial q}{\partial
 P_n}\;\dot{P}_n(t)\;=\;\frac{\partial \Delta {\cal H}(q,\,p)}{\partial
 p}\;\;\;.
 \label{12}
 \ea
So the canonicity demand is incompatible with osculation. In other
words, whenever a momentum-dependent perturbation is present, we
still can use the ansatz (\ref{5}) for calculation of the
coordinates and momenta, but can no longer use (\ref{14}) for
calculating the velocities. Instead, we must use (\ref{13}).
Application of this machinery to the case of orbital motion is
depicted on Fig.1. Here the constants $\;C_j\,=\,(Q_n,\,P_n)\;$
parametrise instantaneous ellipses which, for nonzero $\;\Phi\;$,
are \textbf{not} tangent to the trajectory. (For more details see
Efroimsky \& Goldreich (2003).) In the case of orbital motion, the
situation will be similar, except that, instead of the
instantaneous Keplerian conics, one will deal with instantaneous
Eulerian cones (i.e., with the loci of the rotational axis,
corresponding to non-perturbed spin states).

\subsection{Orbital motion}

In the orbital-motion case, osculation means the following. Let
the unperturbed position be given, in some fixed Cartesian frame,
by vector function $\,\efbold\,$:
 \ba
% \nonumber
   \erbold\;=\;\efbold\left(C_1\,,\;.\,.\,.\,,\;C_6\,\,,\;t
 \right)\;\;\;,\;\;\;\;\;\erbold\;\equiv\;\{\,x\,,\;y\,,\;z
 \,\}\;\;\;.
 \label{13}
 \ea
Employing this functional ansatz also under disturbance, we get
the perturbed velocity as
 \ba
% \nonumber
 \doterbold\;=\;{\bf\vec g}\left(C_1(t)\,,\;.\,.\,.\,,\;C_6(t)\,\,,\;t
 \right)\;+\;\Phibold\left(C_1(t)\,,\;.\,.\,.\,,\;C_6(t)\,\,,\;t
 \right)~~~~~~~~~~~~~~~~~~~
 \label{14}
 \ea
 where
 \ba
 {\bf\vec g}\;\equiv\;\frac{\partial \efbold}{\partial
 t}\;\;\;\;\;\mbox{and}\;\;\;\;\;\;\Phibold\;\equiv\;\sum_{j=1}^{\infty}\,\frac{\partial \efbold}{\partial
 C_j}\;\dot{C}_j\;\;.~~~~
 \label{15}
 \ea
 The osculation condition is a convenient (but totally arbitrary!) demand that the perturbed
 velocity $\,\doterbold\,$ has the same functional dependence upon
 $\,t\,$ and $\,C_j\,$ as the unperturbed velocity $\,{\bf\vec g}\,$:
 \ba
 \nonumber
 \erbold\left(C_1(t)\,,\;.\,.\,.\,,\;C_6(t)\,\,,\;t
 \right)\;=\;\efbold\left(C_1(t)\,,\;.\,.\,.\,,\;C_6(t)\,\,,\;t
 \right)\;\;\;,\\
 \label{16}\\
 \nonumber
 \doterbold\left(C_1(t)\,,\;.\,.\,.\,,\;C_6(t)\,\,,\;t
 \right)\;=\;{\bf\vec g}\left(C_1(t)\,,\;.\,.\,.\,,\;C_6(t)\,\,,\;t
 \right)\;\;\;.\,
 \ea
or, equivalently, that the so-called Lagrange constraint is
satisfied:
 \ba
 \sum_{j=1}^{\infty}\,\frac{\partial \efbold}{\partial
 C_j}\;\dot{C}_j\;=\;\Phibold\left(C_1(t)\,,\;.\,.\,.\,,\;C_6(t)\,\,,\;t
 \right)\;\;\;{\mbox{where}}\;\;\;\Phibold\;=\;0\;\;\;.
 \label{17}
 \ea
Fulfilment of these expectations, however, should in no way be
taken for granted, because the Lagrange constraint (\ref{17}) and
the canonicity demand (\ref{7} - \ref{8}) are now two independent
conditions whose compatibility is not guaranteed. As shown in
Efroimsky (2002a,b), this problem has gauge freedom, which means
that any arbitrary choice of the gauge function
$\;\Phibold\left(C_1(t)\,,\;.\,.\,.\,,\;C_6(t)\,\,,\;t
 \right)\;$ will render, after substitution into (\ref{13} -
 \ref{14}), the same values for $\;\erbold\;$ and
 $\;\doterbold\;$ as were rendered by Lagrange's choice (\ref{17}).\footnote{~
 Physically, this simply means, $\,\doterbold\,$ on Fig.1 can be decomposed
into
 $\,\bf\vec g\,$ and $\,\Phibold\,$ in a continuous variety of
 ways. Mathematically, this freedom reflects a more general construction that
 emerges in the ODE theory. (Newman \& Efroimsky 2003)}
 As can be seen from (\ref{12}), \textbf{the assumption, that the ``constants" $\,Q\,$ and
$\,P\,$ are canonical, fixes the non-Lagrange gauge}
 \ba
 \sum_{j=1}^{\infty}\,\frac{\partial \efbold}{\partial
 C_j}\;\dot{C}_j\;=\;\Phibold\left(C_1(t)\,,\;.\,.\,.\,,\;C_6(t)\,\,,\;t
 \right)\;\;\;{\mbox{where}}\;\;\;\Phibold\;=\;\frac{\partial \Delta \cal H}{\partial \pbold}\;\;\;.
 \label{22}
 \ea
It is easy to show (Efroimsky \& Goldreich 2003; Efroimsky 2005b)
that this same non-Lagrange gauge simultaneously guarantees
fulfilment of the momentum-osculation condition:
 \ba
 \nonumber
 \erbold\left(C_1(t)\,,\;.\,.\,.\,,\;C_6(t)\,\,,\;t
 \right)\;=\;\efbold\left(C_1(t)\,,\;.\,.\,.\,,\;C_6(t)\,\,,\;t
 \right)\;\;\;,\\
 \label{23}\\
 \nonumber
 \pbold\left(C_1(t)\,,\;.\,.\,.\,,\;C_6(t)\,\,,\;t
 \right)\;=\;{\bf\vec g}\left(C_1(t)\,,\;.\,.\,.\,,\;C_6(t)\,\,,\;t
 \right)\;\;\;.\,
 \ea
Any gauge different from (\ref{22}), will prohibit the canonicity
of the elements. In particular, for momentum-dependent $\;\Delta
\cal H\;$, the choice of osculation condition $\;\Phibold\;=\;0\;$
would violate canonicity.

For example, an attempt of a Hamiltonian description of orbits
about a precessing oblate primary will bring up the following
predicament. On the one hand, it is most natural and convenient to
define the Delaunay elements in a co-precessing (equatorial)
coordinate system. On the other hand, these elements will not be
osculating in the frame wherein they were introduced, and
therefore their physical interpretation will be difficult, if at
all possible. Indeed, instantaneous ellipses on Fig.1 may cross
the trajectory at whatever angles (and may be even perpendicular
thereto). Thence, their orbital elements will not describe the
real orientation or shape of the physical trajectory (Efroimsky \&
Goldreich 2004; Efroimsky 2005a).

 For the first time, non-osculating elements obeying (\ref{26})
 implicitly emerged in (Goldreich 1965) and then in Brumberg et
 al (1971), though their exact definition in terms of gauge
 freedom was not yet known at that time. Both authors noticed
 that these elements were not osculating. Brumberg (1992) called
 them ``contact elements." The osculating
and contact variables coincide when the disturbance is
velocity-independent. Otherwise, they differ already in the first
order of the velocity-dependent perturbation. Luckily, in some
situations
% in orbital dynamics
their secular parts differ only in the second order (Efroimsky
2005a), a fortunate circumstance anticipated yet by Goldreich
(1965).

\subsection{Rotational motion}

In rotational dynamics, the situation of an axially symmetric
unsupported top at each instant of time is fully defined by the
three Euler angles $\,q_n\,=\,\theta\,,\,\phi\,,\,\psi\,$ and
their derivatives
$\,\dot{q}_n\,=\,\dot{\theta}\,,\,\dot{\phi}\,,\,\dot{\psi}$. The
time dependence of these six quantities can be calculated from
three dynamical equations of the second order and will, therefore,
depend upon the time and six integration constants:
 \ba
 \nonumber
 q_n\;=\;f_n\left(S_1\,,\;.\,.\,.\,,\;S_6\,\,,\;t
 \right)\;\;\;,\\
 \label{24}\\
 \nonumber
 \dot{q}_n\;=\;\mbox{g}_n\left(S_1\,,\;.\,.\,.\,,\;S_6\,\,,\;t
 \right)\;\;\;,\;
 \ea
the functions $\,\mbox{g}_n\,$ and $\,f_n\,$ being interconnected
via $\,\mbox{g}_n\,\equiv\,{\partial f_n}/{\partial t}\,$, for
$\,n\,=\,1\,,\,2\,,\,3\,=\,\psi\,,\,\theta\,,\,\phi$.

Under disturbance, the motion will be altered:
  \ba
 \nonumber
 q_n\;=\;f_n\left(S_1(t)\,,\;.\,.\,.\,,\;S_6(t)\,\,,\;t
 \right)\;\;\;,~~~~~~~~~~~~~~~~~~~~~~~~~~~~~~~~~~~~~~\\
 \label{25}\\
 \nonumber
 \dot{q}_n\;=\;\mbox{g}_n\left(S_1(t)\,,\;.\,.\,.\,,\;S_6(t)\,\,,\;t
 \right)\;+\;\Phi_n\left(S_1(t)\,,\;.\,.\,.\,,\;S_6(t)\,\,,\;t
 \right)\;\;\;,\;
 \ea
where
 \ba
\Phi_n\left(S_1(t)\,,\;.\,.\,.\,,\;S_6(t)\,\,,\;t
 \right)\;\equiv\;\sum_{j=1}^{6}\frac{\partial f_n}{\partial
 S_j}\;\dot{S}_j\;\;\;.
 \label{26}
 \ea
Now choose the ``constants" $\,S_j\,$ to make canonical pairs
$\,(Q,\,P)\,$ obeying (\ref{7} - \ref{8}), with $\,{\cal H}^*\,$
being zero for $\,(Q,\,P)\,=\,(L_o\,,\,G\,,\,H\,,\,{\it
l}_o\,,\,g_o\,,\,h)\,$. Then insertion of (\ref{7}) into
(\ref{26}) will result in
 \ba
\Phi_n\left(S_1(t)\,,\;.\,.\,.\,,\;S_6(t)\,\,,\;t
 \right)\;\equiv\;\sum \frac{\partial f_n}{\partial
 Q}\;\dot{Q}\;+\;\sum \frac{\partial f_n}{\partial
 P}\;\dot{P}\;=\;\frac{\partial \Delta{\cal H}(q,\,p)}{\partial p_n}\;\;\;,
 \label{27}
 \ea
so that the canonicity demand (\ref{7} - \ref{8}) violates the
gauge freedom in a non-Lagrange fashion. This is merely a
particular case of (\ref{12}).

This yields two consequences. One is that, in the canonical
formalism, calculation of the angular velocities via the elements
must be performed not through the second equation of (\ref{24}) but
through the second equation of (\ref{25}), with (\ref{27})
substituted therein. This means, for example, that in Kinoshita
(1977) expressions (2.6) and (6.24 - 6.27) render not the angular
velocity of the Earth relative to the precessing frame (wherein the
Andoyer variables were defined) but the angular velocity relative to
an inertial frame. (For an extended explanation of this fact see
Efroimsky 2005b.) This, however, is not a drawback of the
Kinoshita-Souchay theory but rather its advantage, because it is the
angular velocity relative to an inertial frame that is directly
measurable. (Schreiber et al 2004)

The second consequence is that, if we wish to make our Andoyer
variables osculating (so that the second equation of (\ref{24})
could be used), the price to be payed for this repair will be the
loss of canonicity. (Angular-velocity-dependent disturbances
cannot be accounted for by merely amending the Hamiltonian!) The
osculating elements will obey non-canonical dynamical equations.

To draw to a close, we would add that, under some special
circumstances, the secular parts of contact elements may coincide
in the first order with those of their osculating
counterparts.\footnote{~In regard to orbital motions, this
possibility was anticipated yet in 1965 by Peter Goldreich. As
demonstrated by Efroimsky (2005a), this is true for constant rate
of frame precession (but not for variable precession).} Whether
this will be the case for the Earth or Mars remains to be
investigated. This matter will be crucial for examining the
validity of the presently available computations of the history of
Mars' obliquity.\footnote{~The pioneer study on this topic was
conducted by Ward (1973) in a direct manner and was, therefore,
exempt from the problems associated with the loss of osculation.
However, some of his successors chose the canonical formalism and
exploited the Hamiltonian borrowed from the Kinoshita theory.}

\section{Conclusions}

In this talk we have explained why the Hamiltonian theory of Earth
rotation renders non-osculating Andoyer elements. We have also
explained how this defect of the theory should be mended.

In attitude mechanics, osculation loss has the same consequences
as in the theory of orbits: while this defect of the theory has no
influence upon the theory's predictions for the figure axis of the
planet, it considerably effects the predictions for the
orientation of the instantaneous axis of rotation.

In our paper Efroimsky (2005b) we shall demonstrate that, even
though the Andoyer variables in the Kinoshita-Souchay theory are
introduced in the precessing frame of the Earth orbit, they return
the angular velocity not relative to that frame, but relative to an
inertial one. This is not a drawback of this theory but rather its
advantage, because it is the angular velocity relative to an
inertial frame that is directly measurable at present. (Schreiber et
al 2004)

\end{document}